# High-efficiency and full-space manipulation of electromagnetic wave-fronts with metasurfaces

Tong Cai[1, 2], GuangMing Wang[2], ShiWei Tang[3], HeXiu Xu[1, 2], JingWen Duan[4], HuiJie Guo[1], FuXin Guan[1], ShuLin Sun[4], Qiong He[1, 5, *] and Lei Zhou[1, 5, *]

[1]*State Key Laboratory of Surface Physics, Key Laboratory of Micro and Nano Photonic Structures (Ministry of Education), and Department of Physics, Fudan University, Shanghai 200433, China*

[2]*Air and Missile Defend College, Air force Engineering University, Xi' an, 710051, China*

[3]*Department of Physics, Faculty of Science, Ningbo University, Ningbo, 315211, China*

[4]*Shanghai Engineering Research Center of Ultra-Precision Optical Manufacturing, Green Photonics and Department of Optical Science and Engineering, Fudan University, Shanghai 200433, China*

[5]*Collaborative Innovation Center of Advanced Microstructures, Nanjing, 210093, China.*

*Corresponding Authors: Qiong He, E-Mail: qionghe@fudan.edu.cn;  Lei Zhou, E-mail: phzhou@fudan.edu.cn

# Abstract

Metasurfaces offered great opportunities to control electromagnetic (EM) waves, but currently available meta-devices typically work either in *pure reflection* or *pure transmission* mode, leaving half of EM space completely unexplored. Here, we propose a new type of metasurface, composed by specifically designed meta-atoms with polarization-dependent transmission and reflection properties, to efficiently manipulate EM waves in the full space. As a proof of concept, three microwave meta-devices are designed, fabricated and experimentally characterized. The first two can bend or focus EM waves at different sides (i.e., transmission/reflection sides) of the metasurfaces depending on the incident polarization, while the third one changes from a wave bender for reflected wave to a focusing lens for transmitted wave as the excitation polarization is rotated, with all these functionalities exhibiting very high efficiencies (in the range of 85%-91%) and total thickness $\sim \lambda/8$. Our findings significantly expand the capabilities of metasurfaces in controlling EM waves, and can stimulate high-performance multi-functional meta-devices facing more challenging and diversified application demands.

# I. Introduction

As the basis of nearly all optical devices, manipulating electromagnetic (EM) wave-front as desired is crucial in modern photonic research. Natural materials exhibit limited variation ranges of permittivity and permeability, so that EM devices made by them are typically too bulky in size and of restricted functionalities, both being unfavorable for EM integration [1-3]. Recently, metamaterials (artificial materials made by subwavelength microstructures with tailored EM properties) and particularly their planar version, metasurfaces, have demonstrated strong capabilities to manipulate EM waves, generating many fascinating effects such as anomalous refraction/reflection [4-6], propagating wave to surface wave coupling [7-10], planar holograms [11-13], focusing lens [14-18], photonic spin Hall effect [19-21], and many others [22-37]. These powerful wave-manipulation abilities have led to many metasurface-based functional EM devices [14, 25, 28, 38-41], which are usually thin, flat, and exhibit diversified and multiple functionalities, all being very promising for modern integration-optics applications.

Despite of the great successes so far achieved, we note that the wave-manipulation capabilities of metasurfaces are far less explored, which also limit the application potentials of them. For example, high-efficiency metasurfaces usually work either in pure reflection mode [4-6, 20-25] (see Fig. 1(a)) or pure transmission mode [25-28, 34, 35, 39], which means that they can only efficiently manipulate either the reflected wave-front or transmitted one, leaving half of EM space totally unutilized. While some metasurfaces could in principle control the wave-fronts of both transmitted and reflected waves [5, 6, 36], the phase gradients provided for transmitted and reflected waves are usually identical leading to locked wave-manipulation functionalities at two sides of the metasurfaces, not mentioning the low efficiencies of these devices due to the undesired multi-mode generations (see Fig. 1(b)). Although many multi-functional metasurface-based devices have been proposed, they typically integrate different functionalities working for either transmission or reflection mode [25, 28], not for both. It is highly desired to expand the wave-manipulation capabilities of metasurfaces to the full EM space, offering the metasurfaces independently controlled functionalities at their two different sides.

In this paper, we propose a new strategy to design metasurfaces that can manipulate the wave-fronts of EM waves in *full* space and with very *high efficiencies* (see Figs. 1(c-d)). The key step is to design a collection of meta-atoms which are perfectly transparent or reflective for incident waves polarized along two orthogonal directions, yet exhibiting tailored (transmission/reflection) phases covering the whole 360° range. We can thus utilize these set of meta-atoms to construct metasurfaces to efficiently and independently control the wave-fronts of EM waves at different sides of the metasurfaces, dictated by the incident polarization. As a proof of concept, we experimentally realized three microwave meta-surfaces: the first two can bend or focus EM waves at two sides of the metasurfaces while the third one combines the functionalities of wave-bending (for reflected wave) and focusing (for transmitted wave) into one single device. In particular, all these devices exhibit very high working efficiencies for all functionalities (in the range of 85%-91%). Our findings open up new possibilities to realize high-efficiency multi-functional meta-devices working in the full space, which can lead to many exciting applications in different frequency domains.

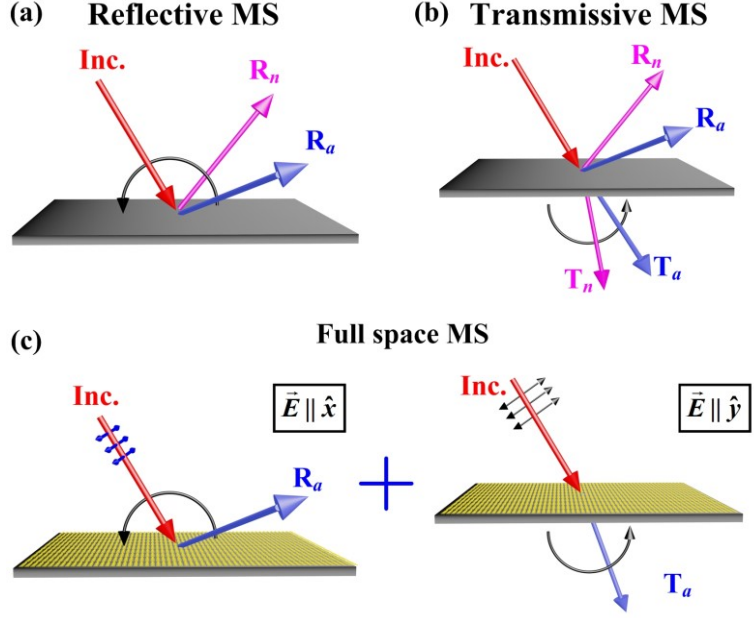

FIG. 1 (color online). Working principle and advantages of the full-space metasurface. Conventional metasurfaces working in (a) reflection or (b) transmission geometries suffer from the issues of restricted working space, low efficiency due to multi-mode generations and locked phase gradients for transmitted and reflected waves. (c) The newly proposed full-space metasurface can efficiently manipulate electromagnetic wave-fronts at both sides of the device with independent functionalities, triggered by incident waves with different polarizations.

## II. Concept and meta-atom design

We describe our strategy to realize the full-space wave-front control, starting from discussing how to design appropriate meta-atoms. Consider a meta-atom exhibiting mirror symmetry, then its EM characteristics can be described by two diagonal Jones matrices $R = \begin{pmatrix} r_{xx} & 0 \\ 0 & r_{yy} \end{pmatrix}$ and $T = \begin{pmatrix} t_{xx} & 0 \\ 0 & t_{yy} \end{pmatrix}$ with $r_{xx}, r_{yy}, t_{xx}$ and $t_{yy}$ denoting the reflection/transmission coefficients for waves polarized along two principle axes $\hat{x}$ and $\hat{y}$. In lossless systems, we have $|r_{xx}|^2 + |t_{xx}|^2 = 1$ and $|r_{yy}|^2 + |t_{yy}|^2 = 1$ due to the energy conservation. To achieve independent yet highly efficient controls on both transmitted and reflected waves, we require our meta-atoms to be perfectly reflective for $\hat{x}$-polarized incident wave (i.e., $|t_{xx}| = 0, |r_{xx}| = 1$), and perfectly transparent for $\hat{y}$-polarized incident wave (i.e., $|t_{yy}|=1, |r_{yy}|=0$). Moreoever, the phases associated with their two non-zero coefficients $r_{xx}$ and $t_{yy}$, denoted by $\varphi^r_{xx}$ and $\varphi^t_{yy}$, respectively, can be freely tuned by varying the structural details of the meta-atoms. If these meta-atoms can be designed, we can thus ultilize them to construct a metasurface exhibiting the desired phase distributions (i.e., $\varphi^r_{xx}(x,y)$ and $\varphi^t_{yy}(x,y)$) to realize certain pre-determined functionalities for controlling reflected and transmitted wave-fronts, under $\hat{x}$- and $\hat{y}$-polarized excitations, respectively.

Figure 2(a) shows the designed meta-atom that can exhibit the mentioned polarization-dependent EM charateristics. As shown in the inset to Fig. 2(a), the meta-atom consists of four metallic layers seperated by three 1.5 mm-thick F4B dietric spacers (with $\varepsilon_r = 2.65 + 0.01i$). The first two layers are ansiotropic metallic crosses consisting of two metallic bars with lengths carefully adjusted, while the bottom two layers are topologically different which are *contrinuous* metallic stripes (along *x* direction) decrorated with small perpendicular metallic bars (along *y* direction). Now the advantages of design are clear. Those *x*-orientated *continuous* metallic stripes at the bottom two layers essentially work as an *effective optical grating* to efficiently block the $\hat{x}$-polarized wave and leave only the $\hat{y}$-polarized waves to pass through, in the frequency regime ( around 10 GHz) studied in this paper. The roles of the top two metallic crosses are to further tune the phases and amplitudes of the transmitted and reflected waves, for two different polarizations. Consider first the $\hat{x}$-polarization. The coupling between top metallic resonators with the bottom continuous stripes generates two magnetic resonances, which can dramatically change the reflection phase as a function of frequency. Figure 2(c) depicts the simulated spectra of reflection amplitude and phase of a typical meta-atom (periodically replicated) under $\hat{x}$-polarization excitation. Indeed, within the frequency interval (7-13 GHz) of interest, our meta-atom can nearly totally reflect $\hat{x}$-polarized wave with reflection-phase $\varphi_{xx}^r$ varying from -180° to 180° as frequency passes through the magnetic resonance at 9.1 GHz [8, 22, 23]. In this particular example, only one magnetic resonance appears since the bar in the second layer is also continuous ($d_2 = 11$ mm) which does not generate a resonance. Two magnetic resonances can appear if the bar in the second layer also exhibits a finite length (i.e., $d_2 < 11$ mm). Therefore, we can have an expanded freedom to design our meta-atom by setting both $d_1$ and $d_2$ freely adjustable (see Section *A* of Supporting Information [42]). For the $\hat{y}$-polarization, however, EM waves can only "see" the *y*-orientated metallic bars in each layer. Since all *y*-orientated metallic bars are of short lengths, their Lorentz resonances are at frequencies much higher than the frequency region that we are interested in, and therefore, EM waves of this polarization can easily pass through each layer. The coupling between different layers can further enhance the transmission by forming a series of Fabry-Perot (FP) transmission modes (see Supporting Information *A* [42]). Via adjusting the geometrical parameters, we can appropriately cascade the generated transmission resonances to get a wide-band transparent window with controllable transmission phase. Figure 2(d) shows the simulated spectra of transmission amplitude $|t_{yy}|$ and phase $\varphi_{yy}^t$ for a typical sample, where we find that $|t_{yy}| > 0.84$ within the frequency band 7-13 GHz while the variation range of $\varphi_{yy}^t$ can cover the whole 360° range.

We fabricated a microwave sample (with a size of 330 mm × 330 mm) consisting of a *periodic* array of a typical meta-atom. Figure 2(b) shows the top-view and bottom-view pictures of the sample. We then experimentally characterized its transmission/reflection characteristics, and compared the measured results with the simulated spectra in Figs. 2(c) and 2(d). Excellent agreement is noted between measured and simulated spectra.

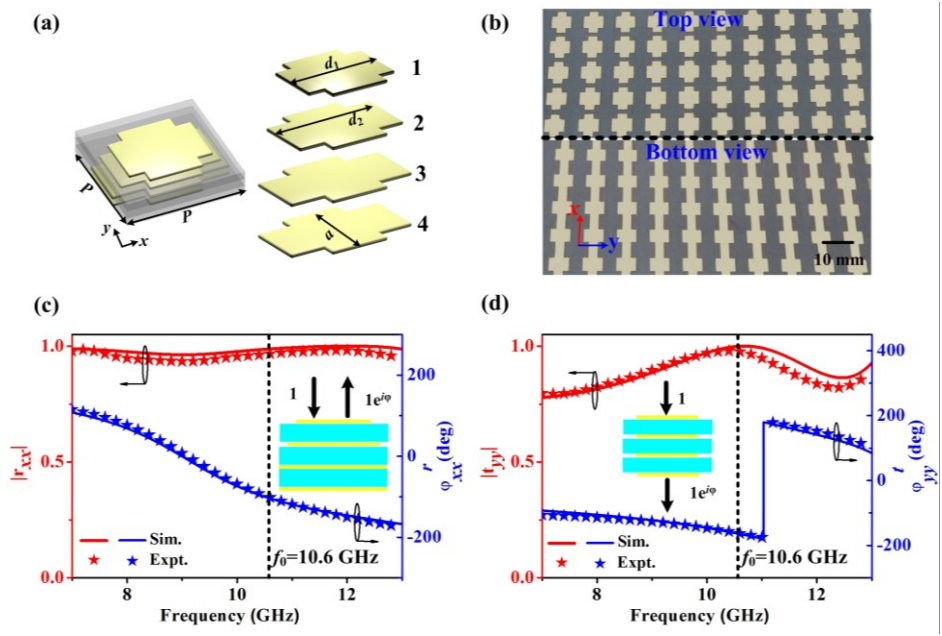

FIG. 2. (color online) Design and characterization of the proposed meta-atom. (a) Schematics of the proposed meta-atom composed by four metallic layers separated by three F4B spacers ($\varepsilon_r = 2.65 + 0.01i$, $h$=1.5 mm). The following geometrical parameters are fixed: width of each *y*-orientated bar is $w_1 = 5$ mm, width of each *x*-orientated bar/stripe is $w_2 = 4$ mm, lengths of the *x*-orientated stripes in the third and fourth layers are fixed to the periodicity $d_3 = d_4 = P = 11$ mm. Other parameters ($d_1, d_2$ and lengths of all y-orientated bars *a*) are tuned appropriately in designing each meta-atom. (b) Top view (left) and bottom view (right) pictures of a fabricated metasurface consisting of a periodic array of meta-atoms with $a = 6.3$ mm, $d_1 = 9$ mm and $d_2 = 11$ mm. Measured and FDTD simulated amplitude/phase spectra of reflection (c) and transmission (d) for the periodic metasurface under excitations with different polarizations.

Figure 2 illustrates that such a meta-atom structure is an ideal building block to construct our meta-surfaces to achieve the full-space wave-front control. Now that the reflection and transmission phases ($\varphi^r_{xx}$ and $\varphi^t_{yy}$) are dictated by the magnetic resonances (for $\hat{x}$-polarization) and the transmission resonances (for $\hat{y}$-polarization), respectively, we understand that changing the geometric structural details of our meta-atom can significantly tune the two phases via varying the corresponding resonance-mode positions. In addition, we understand that these geometrical parameters have different roles in affecting the two phases.

Obviously, structural parameters $d_1, d_2$ are mainly responsible for $\varphi_{xx}^r$ while the parameter $a$ mainly changes $\varphi_{yy}^t$. These nearly delinked control abilities make our realistic design relatively easy, and we can design metasurfaces with arbitrary distributions of $\varphi_{xx}^r(x,y)$ and $\varphi_{yy}^t(x,y)$ according to their desired functionalities, via choosing meta-atoms with carefully adjusted geometrical parameters based on the parameter maps recorded in Supporting Information *B* [42]. We note that in adjusting these geometrical parameters, the reflection/transmission amplitides of the meta-atoms $|r_{xx}(x,y)|$ and $|t_{yy}(x,y)|$ can remain at very high values, which ensure the high-efficiency of the realized functionality (see Supporting Information *B* [42]). We note that our mechanism is obviously different from previous attempts of making multi-functional devices exploring only half of the EM space [12, 14, 15, 21, 24-27], and those attempts utilizing the full EM space but exhibiting locked and low-efficiency functionalities [4, 5]. In the following sections, we will discuss several examples to illustrate our concept.

## III.    Experimental results and discussions

**A. Full-Space Beam Deflector**

As the first example, we employ our meta-atoms to design a high-efficiency beam deflector working in the full space. To achieve this goal, we require that $\varphi_{xx}^r$ and $\varphi_{yy}^t$ exhibit the following distributions

$$\begin{cases} \varphi_{xx}^r = C_0 + \xi_1 \cdot x \\ \varphi_{yy}^t = C_1 + \xi_2 \cdot x \end{cases}, \quad (1)$$

where $C_0$ and $C_1$ are two constants, $\xi_1$ and $\xi_2$ are two phase gradients which determine the bending angles of the anomalously reflected and refracted beams, respectively [4-7]. Set the working frequency as $f_0 = 10.6\, \text{GHz}$, we chose six meta-atoms to form a supercell for our metasurface, and then optimized their geometrical parameters to make Eq. (1) satisfied with $\xi_1 = -0.43k_0$ and $\xi_2 = 0.43k_0$, where $k_0 = 2\pi f_0/c$ with $c$ being the speed of light. The structural details of the optimized meta-atoms are summarized in Section *C* of Supporting Information [42]. To validate our design, we depict in Figs. 3(c-d) the finite-difference-time-domain (FDTD) simulated distributions of reflection/transmission amplitude/phase of the designed meta-surface. We find that the two phase distributions match well with the theoretical curves, meanwhile these meta-atoms exhibit competitive values of transmission/reflection amplitudes ($|r_{xx}|>0.93$ and $|t_{yy}|>0.86$), which can guarantee the high-efficiency operations of our meta-devices.

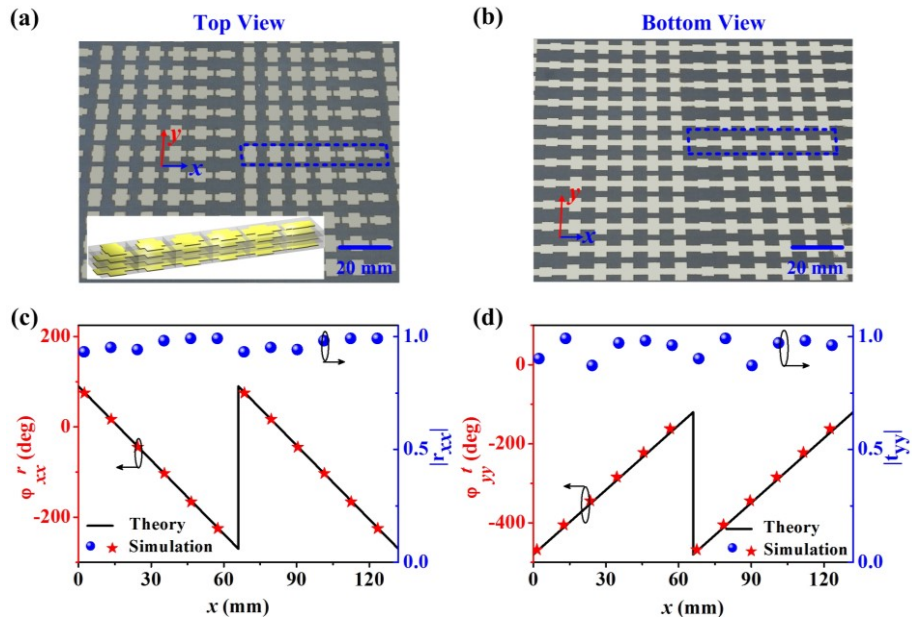

FIG. 3. (color online) Design/fabrication of the full-space beam deflector. (a) Top-view and (b) bottom view pictures of the fabricated full-space beam deflector. FDTD simulated profiles of (c) $\varphi_{xx}^r(x)$, $|r_{xx}(x)|$, (d) $\varphi_{yy}^t$ and $|t_{yy}|$ for the designed full-space beam-deflector, compared with the theoretically requested lines $\varphi_{xx}^r = C_0 - 0.43k_0 \cdot x$ and $\varphi_{yy}^t = C_1 + 0.43k_0 \cdot x$ (black lines). Here the working freqeuncy is $f_0 = 10.6$ GHz.

We fabricated a metasurface sample according to the design (see Figs. 3(a-b) for its top-view and bottom-view pictures), which contains $30\times 30$ meta-atoms with a total size of $330\times 330$ mm$^2$, and then experimentally characterized its wave-manipulation performances. We first characterized the $\hat{x}$-polarization properties of the meta-surface. Shining an $\hat{x}$- polarized microwave normarlly onto our meta-surface, we measured the angular distributions of scattered waves at both reflection and transmission sides of the metasurface. As shown in Figs. 4(a-b), within a frequency interval (10.35 to 11.15 GHz), most input power is reflected to an oblique angle determined by the generalized Snell's law $\theta_r = \sin^{-1}(\xi_1/k_0)$ (solid stars in Fig. 4(a)) [4-7]. The best performance appears at 10.6 GHz where all undesired modes in the full space are suppressed leaving only the anomalous reflection survived. This already implied the good working efficiency of our device. We next characterized the $\hat{y}$-polarization properties of the meta-surface following the same procedures. Figures 4(d) and 4(e) show, respectively, the measured angular power distributions of the scattered waves in reflection and transmission sides for our meta-surface under the illumination of $\hat{y}$-polarized microwaves. As expected, within the frequency interval (10.2 to 11.0 GHz), most input power is now re-directed to the *transmission* chanel at the angle determined by the generalized Snell's law $\theta_t = \sin^{-1}(\xi_2/k_0)$ (solid stars in Fig. 4(d)), with the best performance again appearing at the working frequency 10.6 GHz.

We quantitatively characterized the working efficiencies of our device. At each frequency, we computed the working efficiency as the ratio between the power carried by the desired anomalously deflected mode (either at reflection or transmission side) and a reference value representing the total power carried by the input beam [8, 20, 24]. The former value is obtained by integrating over an appropriate angle region occupied by the desired mode, while the reference is obtained via integrating over the angle region of the specularly reflected mode when the meta-surface is replaced by a metallic plate of the same size (see Section *E* of the Supporting Information for more details [42]). Figs. 4(c) and 4(f) depict the working efficiencies of our device (retrieved from the experimental data shown in Figs. 4(a-b) and 4(d-e)) as functions of frequency, for two polarization-dependent functionalities. We also performed FDTD simulations to compute the angular distributions of the scattering patterns at each frequency for two polarizations (see Supporting Information *D* [42]), from which we successfully obtained the theoretical values of working efficiencies for our device. As shown in Figs. 4(c) and 4(f), the experimentally retrieved efficiencies match well with their corresponding theoretical ones. In particular, our experiments indicate that the working efficiencies of the fabricated device can be as high as 91% (reflection-side) and 85% (transmission-side), while these values are 93% and 88% estimated from FDTD simulations. The slight difference between the measured and simulated results can be attributed to inevitable fabrication errors and imperfections of the incoming wave-fronts generated by our microwave horns.

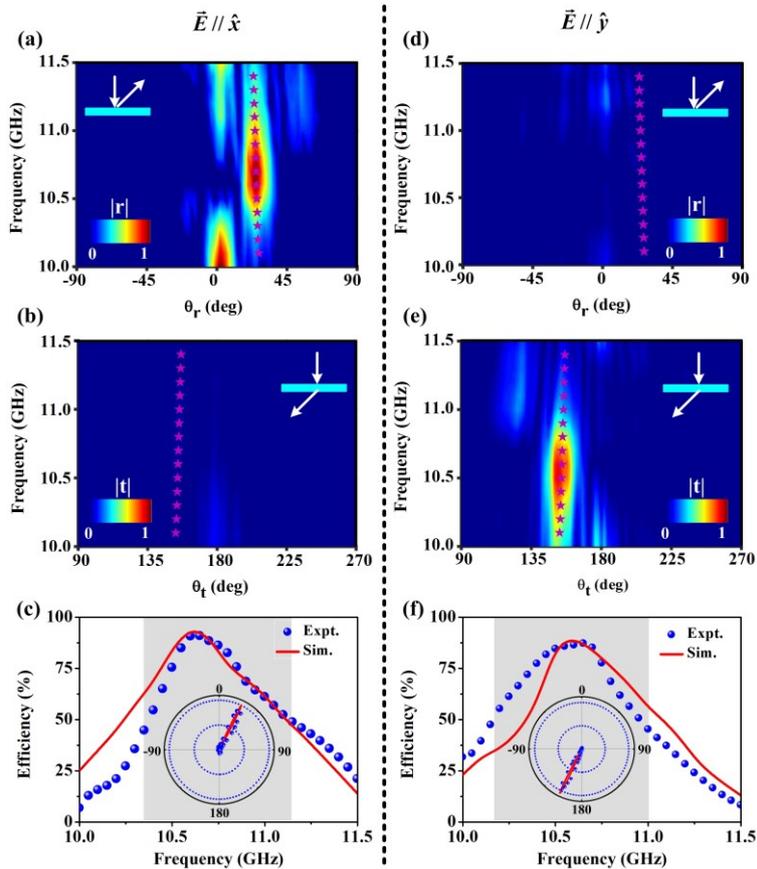

FIG. 4. (color online) Characterizations of the full-space beam deflector under normal incidence. Measured scattered-field intensity (color map) versus frequency and detecting angle at (a) reflection and (b) transmission sides of the metasurface shined by $\hat{x}$-polarized microwaves. (c) Simulated and measured absolute efficiencies of the reflective beam-bending

functionality of the device. Measured scattered-field intensity (color map) versus frequency and detecting angle at (d) reflection and (e) transmission sides of the metasurface shined by $\hat{y}$-polarized microwaves. (f) FDTD simulated and measured absolute efficiencies of the transmissive beam-bending functionality of the device. Insets to (c) and (f) depict the measured (symbols) and simulated scattering patterns of our metasurface illuminated by $\hat{x}$- and $\hat{y}$- polarized waves, respectively, at the frequency $f_0 = 10.6\,\text{GHz}$. All signals are normalized against a reference value obtained by replacing the meta-device with a metallic plate of the same size.

## B. Full-Space Meta-Lens

As another example, we designed a full-space meta-lens, which can focus EM waves at its reflection and transmission sides for $\hat{x}$ and $\hat{y}$-polarized incident waves, respectively (see Figs. 5(a) and 5(e)). To achieve this end, the two phase functions ($\varphi_{xx}^r$ and $\varphi_{yy}^t$) of our meta-device should exhibit the following distributions

$$\begin{cases} \varphi_{xx}^r(x,y) = k_0(\sqrt{F_1^2 + x^2 + y^2} - F_1) \\ \varphi_{yy}^t(x,y) = k_0(\sqrt{F_2^2 + x^2 + y^2} - F_2) \end{cases} \quad (2)$$

where $F_1$ and $F_2$ are two focal lengthes which can be freely selected. Here, still set the working frequency as $f_0 = 10.6\,\text{GHz}$, we designed a full-space meta-lens with two focal lengths $F_1 = F_2 = 80\,\text{mm}$. Different from the deflector realized in the last sub-section, here the meta-lens does not exhibit a super cell, and thus every meta-atom should be carefully optimized such that the two phase distributions can satisify Eq. (2). We fixed the structual details of all meta-atoms adopted based on the parameter maps recorded in Supporting Information $B$ [42], and then fabricated a meta-lens sample according to the design, which contains $14 \times 14$ meta-atoms and has a total size of $154 \times 154\,\text{mm}^2$. Figures 5(b) and 5(f) depict the top-view and bottom-view pictures of the fabricated sample. To validate our design, we calculated the distributions of two relavant phases ($\varphi_{xx}^r$ and $\varphi_{yy}^t$) and amplitudes ($|r_{xx}|$ and $|t_{yy}|$) of our designed/fabricated meta-lens, and depict them in Figs. 5(c, g, d, h), respectively. Clearly, the phase profiles of the fabricated meta-lens follow well with the parabolic distributions dictated by Eq. (2), while the reflection/transmission amplitudes also exhibit high values ($|r_{xx}|>0.92$ and $|t_{yy}|>0.85$), implying the high performances of our device.

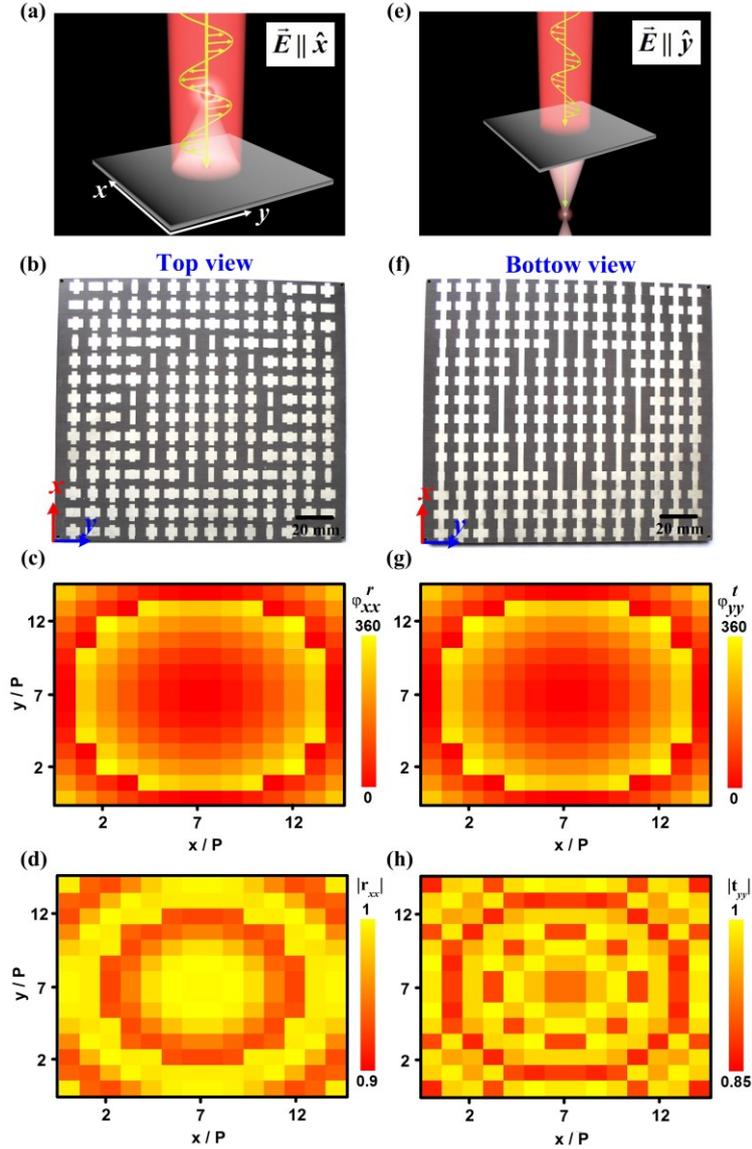

FIG. 5. (color online) Design/fabrication of the full-space meta-lens. Schematic illustration of the performance of our meta-device, which behaves as a (a) reflective lens and (e) a transmissive lens under excitations of $\hat{x}$- and $\hat{y}$- polarized waves, respectively. Top-view (b) and bottom-view (f) pictures of our fabricated sample. FDTD simulated profiles of (c) $\varphi_{xx}^{r}(x,y)$, (d) $|r_{xx}(x,y)|$, (g) $\varphi_{yy}^{t}(x,y)$ and (h) $|t_{yy}(x,y)|$ of the designed/fabricated meta-device. The working frequency is $f_0 = 10.6$ GHz.

With the fabricated sample in hand, we experimentally characterized its full-space focusing performances, with reflection-mode functionality considered first. Shining the sample with an $\hat{x}$-polarized plane wave, we used a monopole antenna (~20 mm long) to measure the electric field distributions at the reflection-side half space. To see clearly the focusing effect, we purposely deducted the incident field from the measured total field, so that the obtained field is solely the scattered one. Figure 6(a) depicts the measured scattered-field distributions on both *xoz* and *yoz* planes at the frequency 10.6 GHz, which are normalized against the maximum value in the pattern. We found that the reflected waves are indeed well converged to a focal point at $z = -77$ mm, identified as the maximum-field point in the $|E_x|^2 \sim z$ curve along the central *z* axis (see Supporting Information *G* [42]). The focal length identified experimentally agrees reasonably with the theoretical value $F_1 = 80$ mm. To check the quality of the focusing effect, we quantitatively evaluated the full-width-at-half-maximum (FWHM) of the focal spot on the focal plane, and found it is approximately 24 mm (see inset to Fig. 6(b)). Obviously, such a value strongly depends on the aperture size of our meta-lens, and can be further reduced by enlarging the total size of our lens. To identify the working bandwidth of our focusing functionality, we show in Fig. 6(b) how the measured and FDTD computed scattered $\vec{L}$-field at the focal point varies against frequency, with the incident $\vec{L}$-field keeping as a constant. The operation bandwidth of our device, defined by the FWHM of the $|E_x|^2 \sim f$ curve, is found as 0.85 GHz (shaded region in Fig. 6(b)). Finally, we used the method reported in refs. [24] and [37] to evaluate the working efficiency of this functionality, defined as the ratio between the powers carried by the focal spot and the incident beam. Our analysis shows that the efficiency is as high as 90.6% (see Supporting Information *H* [42]).

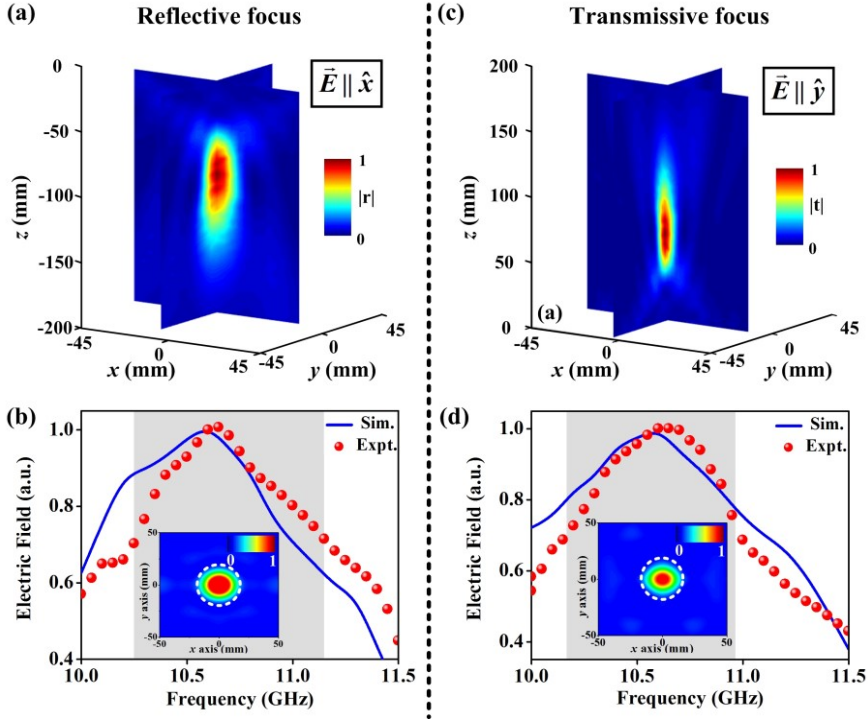

FIG. 6. (color online) Characterizations of the full-space meta-lens. (a) Measured $|E_x|^2$ distributions on both *xoz* and *yoz* planes at the reflection side of the meta-lens under illumination of normally incident $\hat{x}$-polarized wave. (b) Measured/simulated $|E_x|$ at the focal point versus frequency for our meta-lens under $\hat{x}$-polarized excitation. (c) Measured $|E_y|^2$ distributions on both *xoz* and *yoz* planes at the transmission side of the meta-lens under illumination of normally incident $\hat{y}$-polarized wave. (d) Measured/simulated $|E_y|$ at the focal point versus frequency for our meta-lens under $\hat{y}$-polarized excitation. Insets to (b) and (d) depict the measured $|E_x|^2$ and $|E_y|^2$ distributions on the *xy* planes with $z$=-77 mm and $z$=78 mm, respectively, with the dashed-line circles defining the sizes of the focal spots. Here, the working frequency is $f_0 = 10.6$ GHz. All field values are normalized against the maximum value in the corresponding spectrum/pattern.

We then experimentally characterized the focusing performance of our device at the transmission side of the metasurface, which is illuminated by an $\hat{y}$-polarized plane wave. The characterization procedures are essentially the same as those for the reflection-mode functionality, only with the field scanning now carried at the transmission side of the device. The measured $\vec{L}$-field distributions on both *xoz* and *yoz* planes (Fig. 6(c)) at the working frequency clearly revealed the nice focusing effect at the transmission side. The focal length is identified as 78 mm, quite close to the designed value 80 mm (see Supporting Information *G* [42]). The working bandwidth of this functionality is found as 0.75 GHz, indicated as the shaded region in Fig. 6(d) where the spectrum of the $\vec{L}$-field measured at the focal point is shown. We also checked the quality of the focusing effect, and found that the size of the focal spot is about 20 mm (see inset to Fig. 6(d) for the measured field pattern on the focal plane). Finally, we note that this focusing functionality still exhibits high working efficiency 85%, obtained with the same procedure as that for the reflective lens (see Supporting Information *H* [42]).

**C. Full-Space Bifunctional Meta-device**

In previous sections, the metasurfaces that we realized exhibit the *same* functionalities for reflected and transmitted waves. In this section, we further demonstrate that such full-space manipulation is not restricted to realizing identical functionalities. As an illustration, we design a meta-device which combines beam-bending and focusing functionalities in a single device, with such two distinct functionalities working for reflected and transmitted EM waves, respectively (see Figs. 7(a) and 7(d)). To achieve this goal, we require the two phase functions ($\varphi_{xx}^r$ and $\varphi_{yy}^t$) of our meta-device to satisfy the following distributions

$$\begin{cases} \varphi_{xx}^r(x,y) = C_2 + \xi_3 x \\ \varphi_{yy}^t(x,y) = k_0(\sqrt{F_3^2 + x^2 + y^2} - F_3) \end{cases} \quad (3)$$

with $C_2$ being a constant. Here, $\xi_3$ is the phase gradient to determine the angle of the anomalously reflected wave and $F_3$ is the focal length for the meta-lens application working for the transmitted wave, both of which can be freely chosen. In our design, keeping the working frequency still at $f_0 = 10.6$ GHz, we set $\xi_3 = 0.51 k_0$ and $F_3 = 85$ mm without losing generality. Aided by the structual map shown in Section *B* of the Supporting Information [42], we optimized each meta-atom in the metasurface such that the resulting $\varphi_{xx}^r$ and $\varphi_{yy}^t$ distributions satisfy Eq. (3) (see Fig. S13 in Supporting Information *I* [42]), and then fabricated a sample according to the design. Figs. 7(c) and 7(e) depict the top-view and bottom-view pictures of the fabricated sample, which contains 15×15 meta-atoms and has a total size of 165×165 mm$^2$. Again, we emphasize that all optimized meta-atoms exhibit very high values of reflection/transmission amplitudes ($|r_{xx}|$>0.92, $|t_{yy}|$>0.86, see Fig. S13 in Supporting Information *I* [42]), which guarantees that our meta-device must exhibit high working efficiencies.

We first experimentally characterized the beam-bending functionality of the device at the reflection side. Following the experimental procedures described in Section 3.1, we measured the angular distributions of scattered waves in both reflection and transmission sides of the metasurface, which is shined by normally incident $\hat{x}$-polarized EM waves. Fig. 7(c) clealy shows that nearly all incident powers are re-directed to the anomolous-reflection angles dictated by the generalized Snell's laws (pink symbols), within a broad frequency interval (8.4-11.7 GHz). FDTD simulations well reproduced the experimental observations and the results are presented in Section *J* of Supporting Information [42]. Fig. 7(c) already implied that the working bandwidth of this functionality is quite broad, which is reinforced by the experimentally measured efficiency spectrum of the anomolous-reflection as shown in Fig. 7(d). Obviously, the best beam-bending performance is found at about 10.6 GHz with a peak absolute efficiency 88% (simulation result: 92%), and inset to Fig. 7(d) shows that the scattering pattern at this frequency is very clean and contains only one single anomalous reflection mode.

We finally examined the focusing functionality of our device at its transmission side. We experimentally mapped out the $E_y$-field distributions on two high-symmetry planes at the transmission side of our meta-surface, which is shined by a normally incident $\hat{y}$-polarized EM wave at 10.6 GHz. Fig. 7(g) shows clearly that our meta-device now works as a meta-lens that can focus transmitted wave to a focal point, with focal length evaluated as 84 mm, agreeing well with the theoretical design ($F = 85$ mm). The spot size at the focal plane is found as 19 mm (see inset to Fig. 7(h) for the measured field pattern on the focal plane). Utilizing the same approach as in Sec. 3.2, we also experimentally characterized the working efficiency of the meta-lens, and found that the efficiency is roughly 85.2% (see Supporting Information J [42]). The measured $|E_y|$ value at the focal point is depicted in Fig. 7(h) as a function of frequency, from which we identified the working bandwidth of the meta-lens as 0.8 GHz (10.4-11.2 GHz).

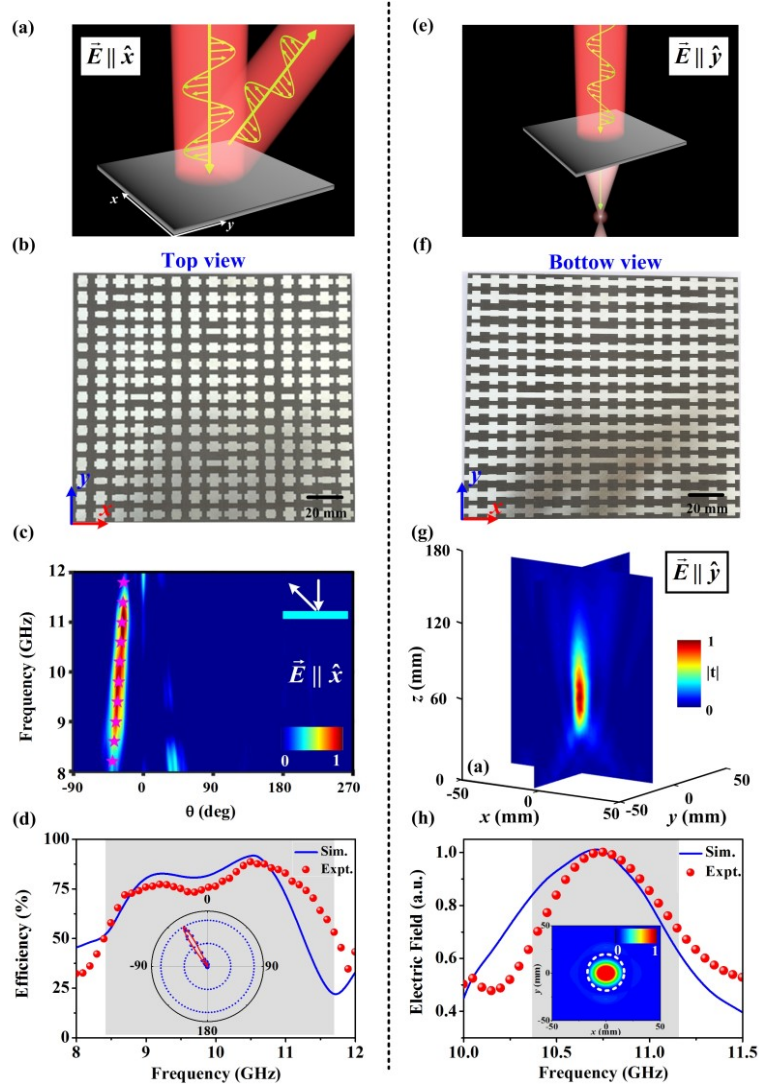

FIG. 7. (color online) Design, fabrication and characterization of a full-space bifunctional meta-device. Schematic illustration of the performance of our meta-device, which behaves as a (a) reflective beam-bender and (e) a transmissive lens under excitations of $\hat{x}$- and $\hat{y}$-polarized waves, respectively. (b) Top-view and (f) bottom-view pictures of our fabricated meta-device. (c) Measured scattered-field intensity (color map) versus frequency and detecting angle at the reflection side of the metasurface shined by normal-incidence wave with $\vec{L}_{\parallel x}$ polarization. (d) Simulated and measured absolute efficiencies of the reflective beam-bending functionality of the device under $\hat{x}$-polarized excitations. Inset shows the measured and FDTD simulated scattering patterns of the meta-surface. (g) Measured $|E_y|^2$ distributions on both *xoz* and *yoz* planes at the transmission side of the metasurface under illumination of normally incident $\hat{y}$-polarized wave. (h) Measured and simulated $\vec{L}$-field amplitude at the focal point as functions of frequency. Inset depicts the measured $|E_y|^2$ distributions on the *xy* plane with $z=84$ mm, with the dashed-line circle defining the size of the focal spot. Here, the working frequency is 10.6 GHz and all field values are normalized against the maximum value inside each spectrum.

## IV. Conclusion

To summarize, we proposed a new type of metasurface that can efficiently control EM wave-fronts in the full space, and experimentally demonstrated the concept in the microwave regime. We designed/fabricated three meta-devices (with total thickness much less than wavelength) and experimentally demonstrated that they can simultaneously realize the beam-bending and/or focusing functionalities in transmission and reflection modes with very high working efficiencies (in the range of 85%-91%), depending on the input polarizations. Our findings open the door to realize functional high-efficiency meta-devices with full-space control abilities in different frequency domains, which are important in modern integration-optics applications.

## V. Acknowledgements

This work was supported by National Basic Research Program of China (No. 2017YFA0303500), National Natural Science Foundation China (Nos. 11474057, 11404063, 11674068, 61372034, 61501499, 11604167), Natural Science Foundation of Shanghai (grant No. 16ZR1445200, 16JC1403100) and Natural Science Foundation of Shaanxi province (Nos. 2016JQ6001).

**Supporting Information**

# High-efficiency and full-space manipulation of electromagnetic wave-fronts with metasurfaces

Tong Cai[1,2], GuangMing Wang[2], ShiWei Tang[3], HeXiu Xu[1,2], JingWen Duan[4], HuiJie Guo[1], FuXin Guan[1], ShuLin Sun[4], Qiong He[1,5,*] and Lei Zhou[1,5,*]

# A. Working mechanism of the meta-atom

Our proposed meta-atom can achieve total reflection and total transmission for incident eletromagnetic (EM) waves with different polarizations. In this subsection, we illustrate the underlying physics via carefully analyzing the EM transmission/reflection chracteristics of the meta-atom under excitations with different polarizations.

Consider first the $\hat{x}$-polarization where the meta-atom should be (nearly) totally reflective but exhibits a controllable reflection phase. As already explained in the main text, our proposed meta-atom consists of 4 metallic layers in which the $x$-oriented metallic bars in the bottom two layers are always continuous stripes but those in the top two layers can exhibit finite lengths. We first consider the case (called meta-atom 1, see Fig. S1a) that only the $x$-oriented metallic bar in the first layer has a finite length ($d_1 = 8$ mm $< P = 11$ mm). Fig. S1c shows the finite-difference-time-domain (FDTD) calculated reflection spectrum of such a meta-atom. A magnetic resonance is generated by the interaction between the top metallic bar and the bottom continuous stripes, evidenced by the reflection dip and the in-phase reflection at 9 GHz and the current distribution depicted in the inset to Fig. S1c. We next consider the second example (called meta-atom 2) where the $x$-orientated bars in the top two layers are both of finite lengths ($d_1$=$d_2$=8.5 mm, see Fig. S1b). The additional freedom provided by the second metallic bar makes such a meta-atom exhibit two magnetic resonances, appearing at 7.2 GHz and 11.9 GHz, respectively (see Fig. S1d). The calculated current distributions (see inset to Fig. S1d) reveal that the low-frequency magnetic resonance is due to the interaction between the upper metallic bars and the lower continuous metallic stripes while the second magnetic resonance is mainly generated by the mutual interaction between two metal bars in the upper two layers. Figure S1 implied that we can easily get any desired reflection phase at a target frequency via tuning the magnetic resonance positions with the reflection amplitude remaining nearly unchanged, for the $\hat{x}$-polarized excitation.

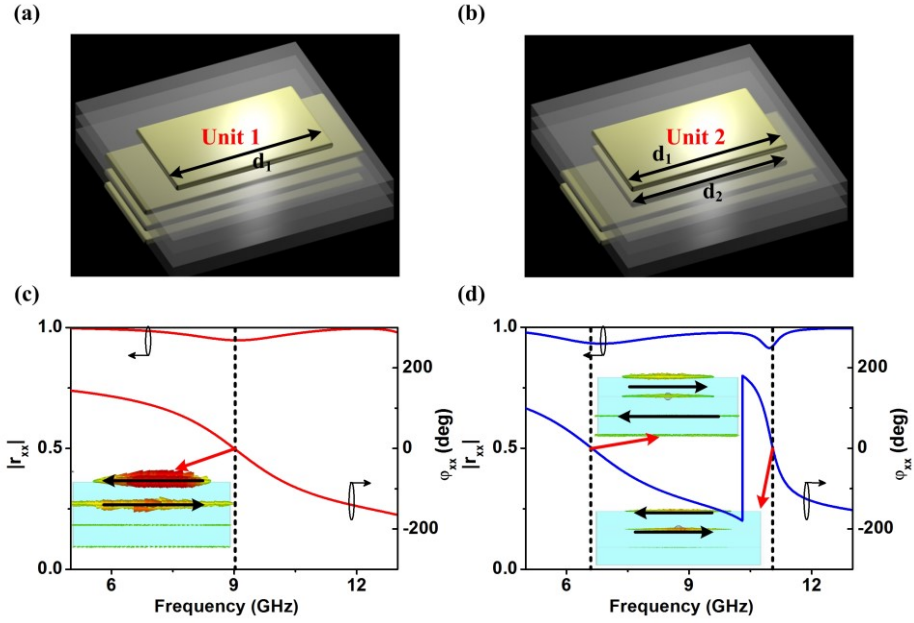

**Figure S1.** Schematics of unit 1 (a) and (b) unit 2. FDTD simulated reflection phase and amplitude for (c) unit 1 with $d_1$=8 mm and (d) unit 2 with $d_2$=$d_1$=8.5 mm. Insets to (c) and (d) show the current distributions at the resonant frequencies.

We now consider the EM properties of the meta-atom under $\hat{y}$-polarization excitation. To illustrate the essential physics, we simplify the complicated structure by only retaining those *y*-oriented bars in each layer since *y*-polarized EM waves can only "see" the *y*-oriented bars under the lowest order approximation. Now the system is a multilayer of metallic patches separated by F4B dielectric spacers. Obviously, there are two topologically different units inside such a multilayer system --- a single metallic patch on a dielectric substrate (called A layer, see Fig. S2(a,b)) and the same metallic patch sandwiched between two dielectric spacers (called A' layer, see Fig. S2(d,e)). We first study the EM characteristics of these two layers. Solid lines in Figs. S2c and S2f depict the FDTD calculated spectra of transmission amplitude and phase for these two layers, from which we find that they can be reasonably described by two effective permittivity $\varepsilon_{eff}^A = 10 - \frac{100^2}{f^2 - 20.98^2}$, $\varepsilon_{eff}^{A'} = 10 - \frac{110^2}{f^2 - 20.8^2}$ at the frequency region that we are interested in, evidenced by the good agreement between FDTD simulations on realistic structures (lines) and transfer-matrix-method (TMM) calculations on effective-medium slabs (symbols).

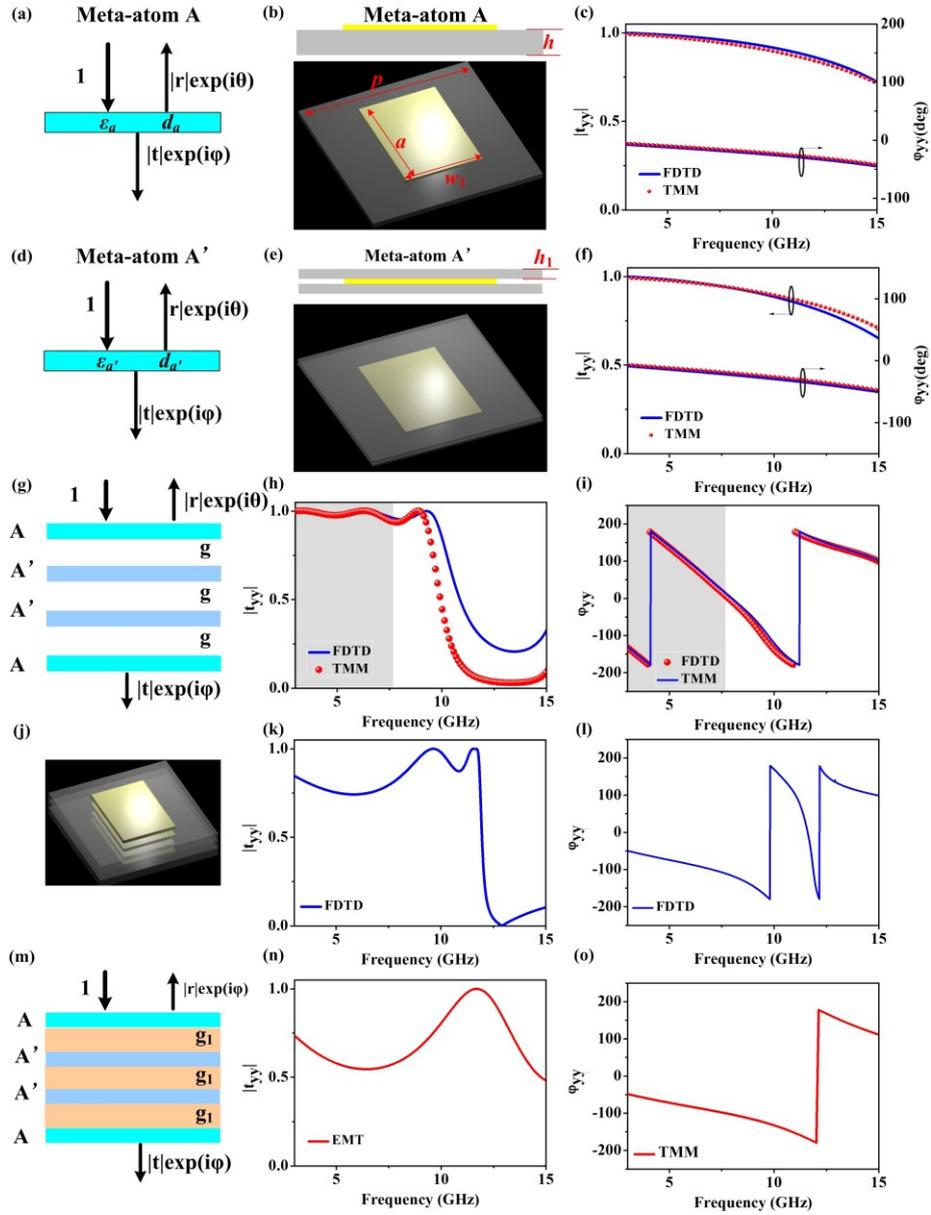

**Figure S2.** (a, d) Effective-medium models and (b, e) schematics of meta-atom A and meta-atom A'. Transmission amplitude and phase spectra for (c) meta-atom A and (f) meta-atom A', calculated by FDTD simulations on realistic structures and TMM calculations on effective-medium slabs. (g) Geometry of a four-layer meta-atom consisting of two layers A and two layers A' separated by air gaps with thickness *g*. Transmission (h) amplitude and (i) phase spectra for such a four-layer meta-atom, calculated by FDTD simulations on realistic structures and TMM calculations on model systems. (j) Schematics of the realistic 4-layer meta-atom in which the gaps separating different layers are F4B dielectric slabs with thickness $g_1$. FDTD simulated (k) transmission amplitude and (l) phase spectra of such a meta-atom; (m) Effective model of the realistic four-layer meta-atom, and its corresponding TMM computed (n) transmission amplitude and (o) phase spectra. The parameters are fixed as *p*=11 mm, *h*=0.1 mm, $w_1$=5 mm, *a*=8 mm, $d_a=d_a$'=0.1 mm, *h*=0.05 mm, *g*=10 mm, $g_1$=1.4 mm.

With the EM properties of individual A and A's layers fully understood, we can then stack them to form a multilayer structure and study the EM characteristics of the resulting multilayer systems. As shown in Fig. S2g, when the air gap between two adjacent layers is large enough, FDTD simulations on realistic multilayer structure are in good agreement with the TMM calculations on the model system in which each realistic layer is replaced by its effective-medium layer (A or A'), both showing that the 4-layer structure can exhibit a series of perfect transmission peaks due to the constructive interferences of multiply scattered waves. Designing the structure to cascade several transmission peaks appropriately, we can find a wide frequency band with high transmission amplitudes and phases covering the $2\pi$ range (see shaded regions in Fig. S2h).

In realistic design, the situation is more complicated, since the spacer between two adjacent layers is much thinner than the case we studied in Fig. S2g and thus the near-field interactions between adjacent layers are more significant. However, the physics remains essentially unchanged. Figure S2k-l depict the FDTD simulated transmission amplitude/phase spectra of a 4-layer meta-atom with spacers in realistic design (gap thickness 1.4 mm). While the FDTD results have quantitative differences with those obtained based on effective medium models (Fig. S2n-o) due to the enhanced near-field interactions between adjacent layers, we note that the essential features of the FDTD spectra are similar to those of the TMM ones.

## B. Design strategy of the full-space metasurfaces

To fasciliate our design, we present in this section detailed parameter maps, based which we can easily fix the structural details of meta-atoms at diferent positions in a metasurface accroding to its desired phase distributions. To this end, we need to calculate how the EM responseses of the meta-atom (e.g., $\varphi_{xx}^r$, $|r_{xx}|$, $\varphi_{yy}^t$, and $|t_{yy}|$) depend on its structural parameters. In principle, we need to present three-dimensional (3D) maps since we have three independent structural parameters (e.g., $a$, $d_1$ and $d_2$). In practice, however, we found it not absolutely necessary to make $d_1$ and $d_2$ fully independent. Instead, we found it is enough to restrict our parameter sorting within two projected parameter spaces: $\{5 < d_1 < 10.8, d_2 = 11\}$ & $\{8 < d_1 = d_2 < 9\}$. Figure S3a-3d illustrate, respectively, how $\varphi_{xx}^r$, $|r_{xx}|$, $\varphi_{yy}^t$, and $|t_{yy}|$ vary against the parameters $a$, $d_1$ and $d_2$, with frequency fixed at $f_0$=10.6 GHz. Obviously, $\varphi_{xx}^r$ is sensitive to $d_1$ and $d_2$ but insensitive to $a$, while $\varphi_{yy}^t$ behaves just oppositely. Changing the structural parameters within the two restricted spaces, we find that the variations of the two phases ($\varphi_{xx}^r$ and $\varphi_{yy}^t$) already cover the whole 360° range, while simultaneously the reflection/transmission amplitudes ($|r_{xx}|$ and $|t_{yy}|$) remain at very high values ($|r_{xx}|>0.92$, $|t_{yy}|>0.84$), which ensures the high working efficiency of the designed metasurface.

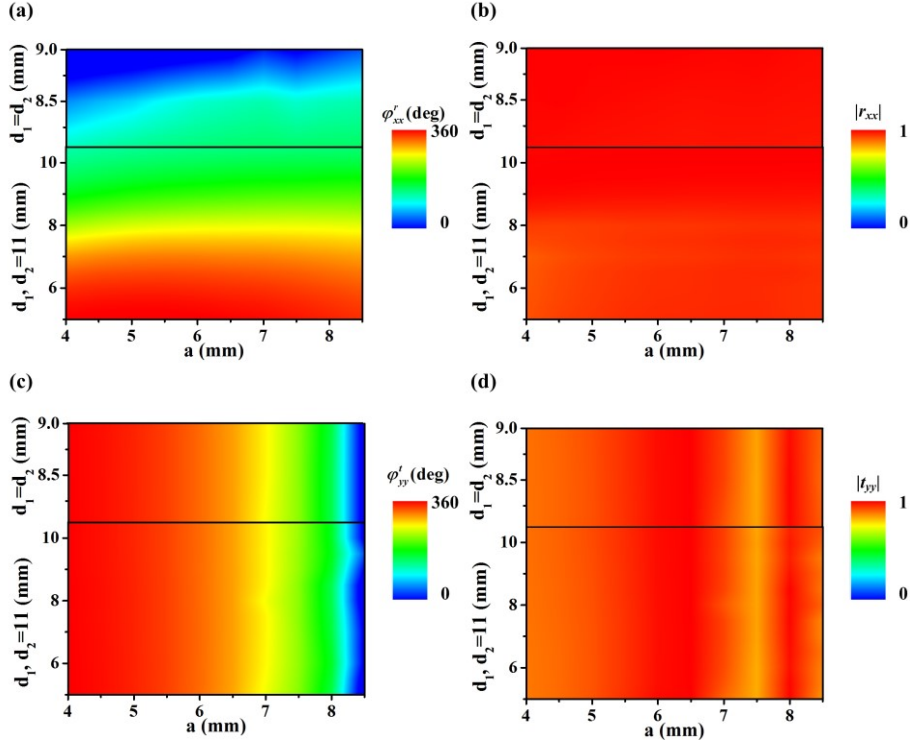

**Figure S3.** (a) Reflection phase $\varphi_{xx}^r$ and (b) the amplitude $|r_{xx}|$ distributions of the meta-atom as functions of $a$ and $d_1$ ($d_2$) for normally incident $\hat{x}$-polarized waves. (c) Transmission phase $\varphi_{yy}^t$ and (d) amplitude $|t_{yy}|$ distributions of the meta-atom as functions of $a$ and $d_1$ ($d_2$) for normally incident $\hat{y}$-polarized waves. Here, the frequency of interest is $f_0$=10.6 GHz.

## C. Detailed structural parameters of three meta-devices designed in this paper

With the structural maps (Fig. S3) in hand, we can then determine the geometrical parameters ($a$, $d_1$ and $d_2$) of all meta-atoms involved in our full-space metasurfaces based on their required phase distributions. Figure S4-S6 depict, respectively, the distributions of meta-atom parameters of three meta-devices realized in this paper.

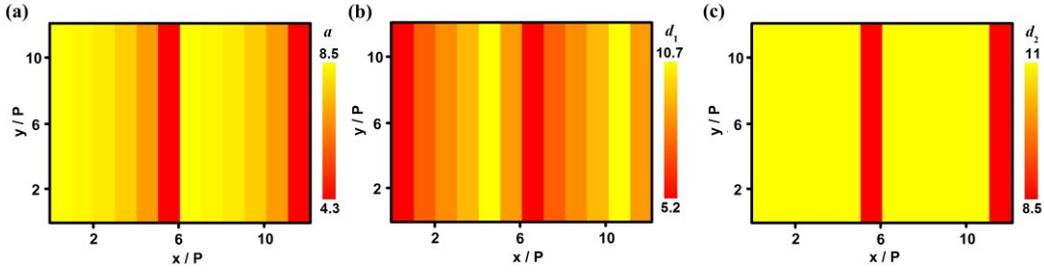

**Figure S4.** Distributions of structural parameters (a) $a$, (b) $d_1$ and (c) $d_2$ of the full-space beam deflector studied in Figs. 3-4 in the main text. All parameters have the unit of mm.

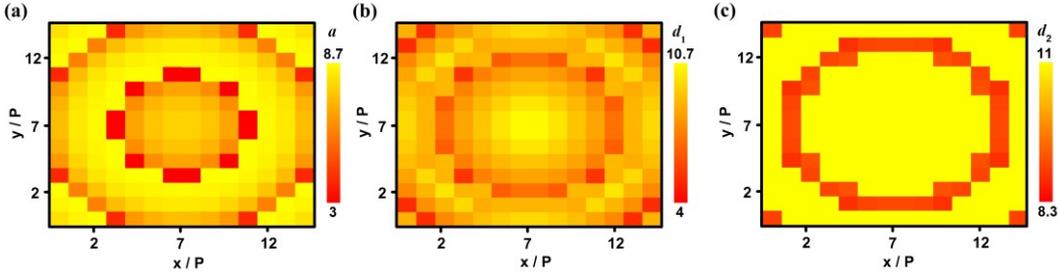

**Figure S5.** Distributions of structural parameters (a) $a$, (b) $d_1$ and (c) $d_2$ of the full-space meta-lens studied in Figs. 5-6 in the main text. All parameters have the unit of mm.

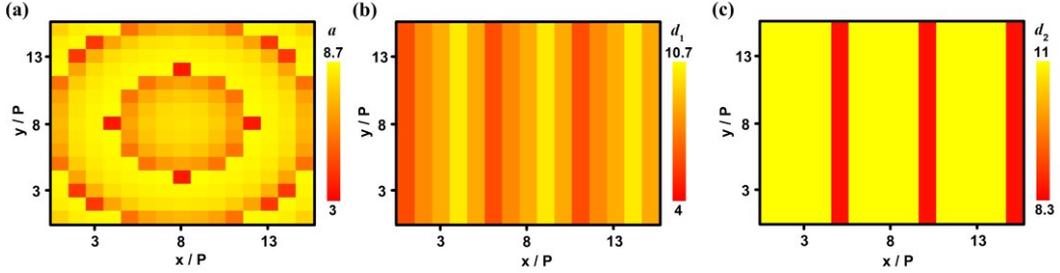

**Figure S6.** Distributions of structural parameters (a) $a$, (b) $d_1$ and (c) $d_2$ of the full-space bifunctional met-device studied in Fig. 7 in the main text. All parameters have the unit of mm.

## D. FDTD simulation results for the full-space deflector

Figure S7 depicts the FDTD simulated scattering properties of the full-space defector under exactly the same excitation conditions as in Fig. 4 of the main text. Obviously, FDTD results are in good agreement with their corresponding experimental results (Fig. 4). In particular, at the working frequency 10.6 GHz, simulations show that the input beams have been redirected to the anomalous-reflection (for $\hat{x}$-polarization excitation) and the anomalous-refraction mode (for $\hat{y}$-polarization excitation) with very high efficiencies, since all undesired modes have been fully suppressed.

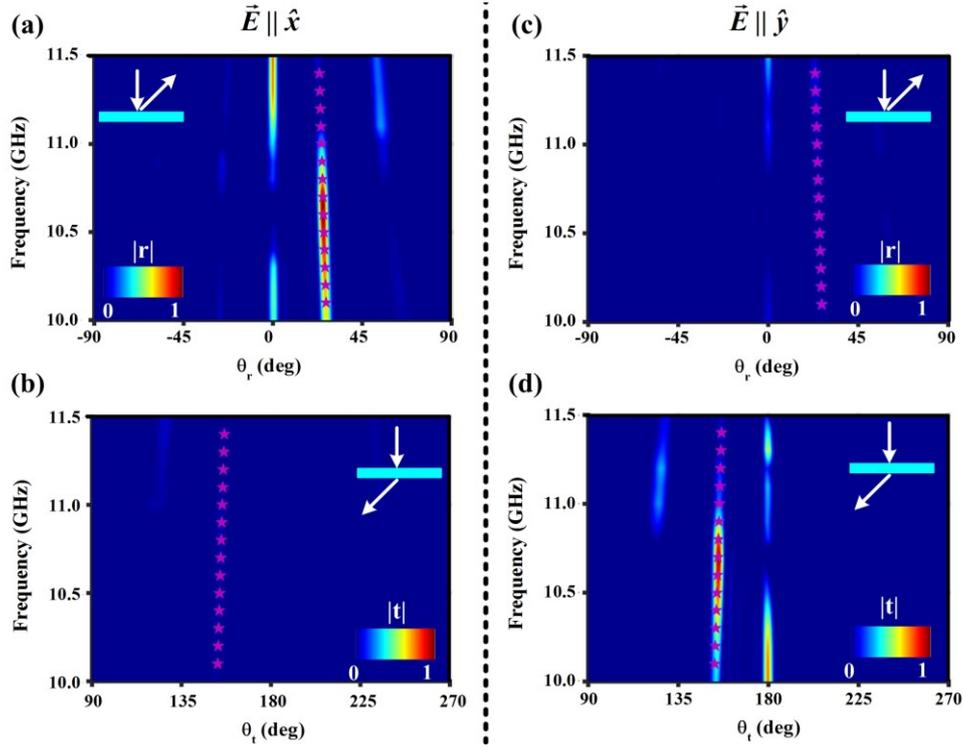

**Figure S7.** FDTD simulated scattered-field intensities (color map) in (a, c) reflection and (b, d) transmission space versus frequency and the detecting angles of our meta-beam-deflector shined by normal-incidence wave with $\vec{E} \parallel \hat{x}$ (a, b) and $\vec{E} \parallel \hat{y}$ (d, e) polarizations, respectively.

## E. Evaluating the working efficiencies of the full-space beam deflector

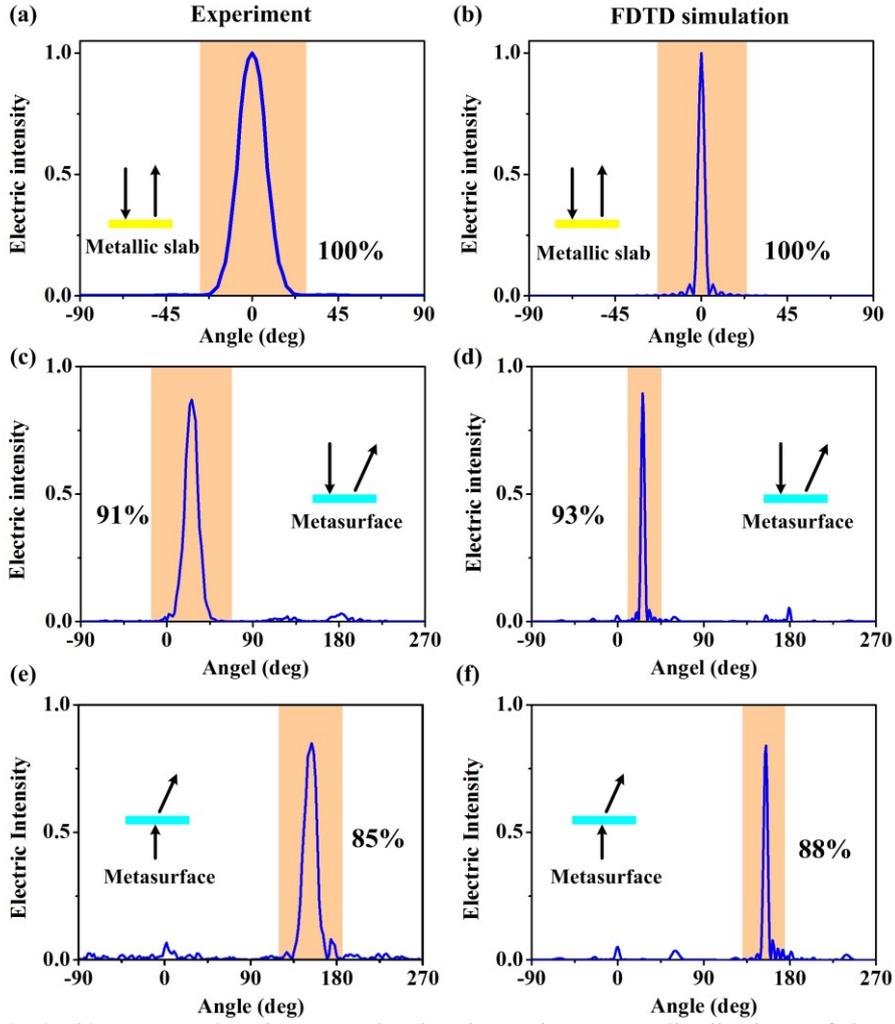

**Figure S8.** (a, b) Measured and FDTD simulated angular power distributions of the scattered field for a metallic plate shined by a normally incident wave. (c, d) Measured and FDTD simulated angular power distributions of the scattered field for our metasurface shined by a normally incident $\hat{x}$-polarized wave. (e, f) Measured and FDTD simulated angular power distributions of the scattered field for our metasurface shined by a normally-incident $\hat{y}$-polarized wave.

In this section, we quantitatively estimate the working efficiencies of our full-space beam deflector based on both experimental data and FDTD results. We define the efficiency as the ratio between the power carried by the deflected beam (anomalously reflected/refracted modes) and that of the incident one. The former is evaluated as the integrated power over the angle ranges occupied by anomalously reflected wave (Fig. S8c, 8d) or the anomalously refracted one (Fig. S8e, 8f), for our metasurface illuminated by normally incident waves with different polarizations at 10.6 GHz, while the latter is obtained by the same technique but with the metasurface replaced by a metal plate of the same size (Fig. S10a, b). The ratio between the two is thus the desired efficiency. Based on this technique, we found that the working efficiency of the anomalous reflection is 91% in experiment and 93% in simulation, and that of the anomalous transmission is 85% in experiment and 88% in simulation.

## F. Incidence-angle-dependent performance of the full-space beam deflector

In this section, we numerically and experimentally illustrate the incidence-angle dependent properties of our full-space beam deflector at the frequency 10.6 GHz.

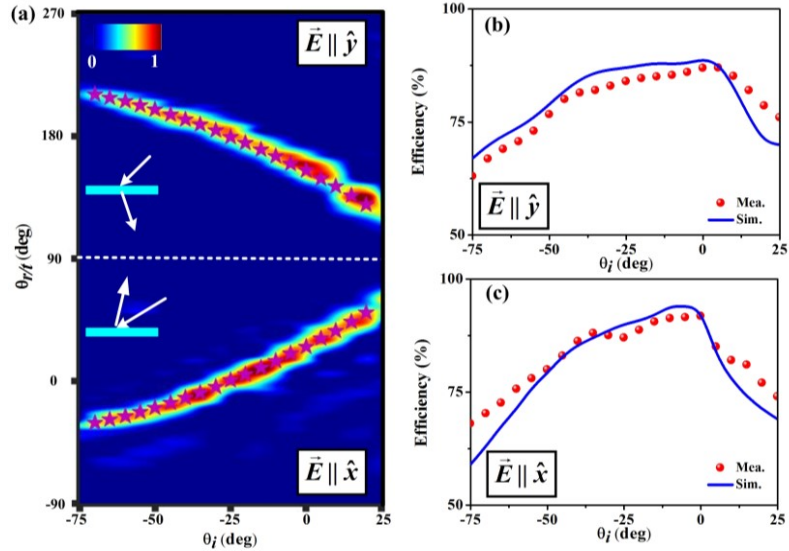

**Figure S9.** Characterizations of our beam deflector under oblique incidence. (a) Measured scattered-field intensities (color map) in reflection (down) and transmission (up) sides versus incidence angles $\theta_i$ and the detecting angles $\theta_r/\theta_t$, for our meta-beam-deflector illuminated by $\hat{y}$-polarized (upper panel) and $\hat{x}$-polarized (lower panel) waves. FDTD simulated and measured absolute efficiencies of (b) anomalous reflection and (c) anomalous refraction of our meta-device under oblique illuminations with $\hat{y}$- and $\hat{x}$-polarizations at 10.6 GHz.

Figure S9a shows how the measured normalized intensities of the anomalous reflection/transmission signals vary against incident ($\theta_i$) and detecting angles ($\theta_r, \theta_t$), for our beam deflector under illuminations of $\hat{x}$- and $\hat{y}$-polarized microwaves. We can see clearly that the relationships between anomalous reflection/transmission angles and the incident angles well satisfy the generalized Snell's law $\theta_{r/t} = \sin^{-1}(\sin\theta_i + \xi/k_0)$ represented by blue stars. Figures S9b and 9c further compare the measured and simulated working efficiencies of our device (using the technique described in Sec. E) for two functionalities as functions of the incidence angle, from which we find the working efficiencies remain at high values. Excellent agreement between measured and simulated results is noted.

## G. Estimations on the focal lengths of the full-space meta-lens

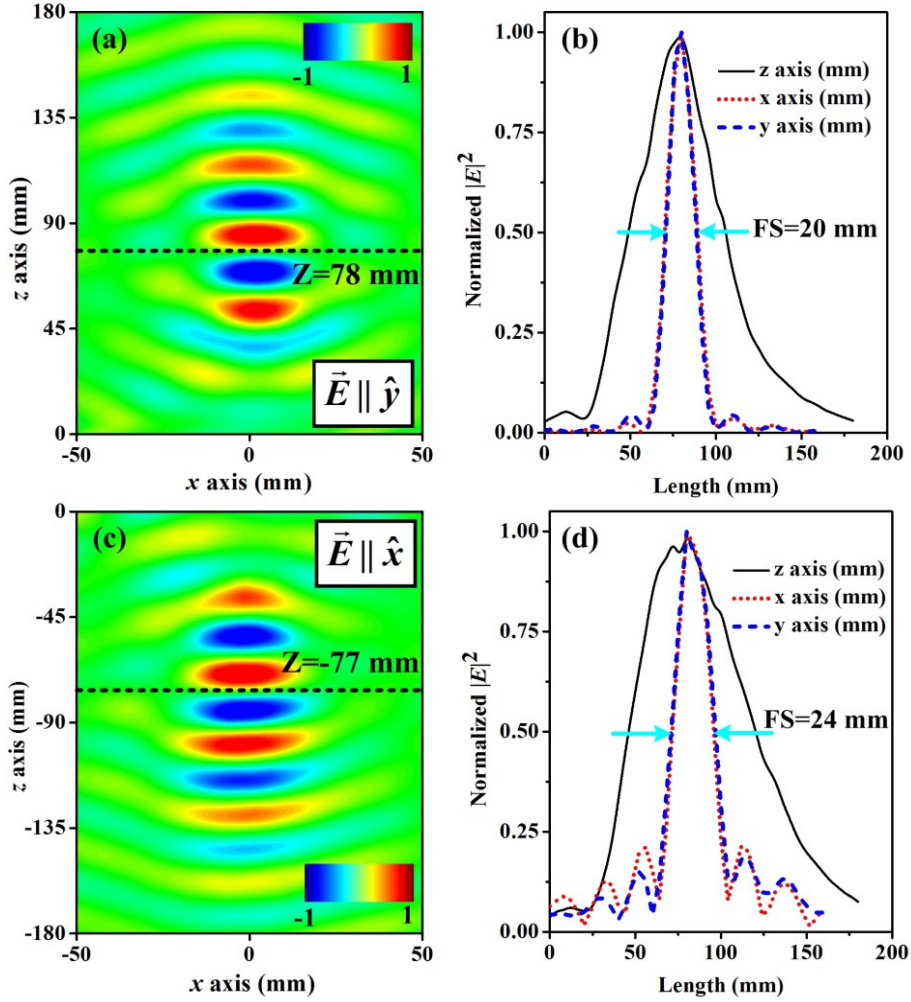

**Figure S10.** Measured $\text{Re}(E_y)$ (a) and $\text{Re}(E_x)$ (c) distributions at *xoz* planes at transmission space and reflection space for our meta-device illuminated by $\hat{y}$- and $\hat{x}$-polarized waves at working frequency of 10.6 GHz, respectively. Calculated (b) $|E_y|^2$ and (d) $|E_x|^2$ distributions along three axes from the measured $\vec{E}$-Field distributions for our meta-lens, respectively.

In this section, we describe how to evaluate the focal lengths of our meta-lenses studied in Figs. 5-6 of the main text. Figure S10a depicts the measured $\text{Re}(E_y)$ distributions on the *xoz* plane at the transmission side of our meta-lens, under the illumination of a $\hat{y}$-polarized plane wave at $f_0 = 10.6$ GHz. Based on such $\text{Re}(E_y)$ distribution, we plot in Fig. S10b (black line) how the measured $|E_y|^2$ varies against *z* on the central *z* axis (with *x*=0), from which we can easily identify the focal point (z=78 mm) where the maximum field value appears. Here, $\text{Re}(E_y)$ contains the phase information, therefore the focal plane is not at the maximum value of $\text{Re}(E_y)$. In order to estimate the spot size on the focal plane, we also measured the $|E_y|^2$ distributions along *x* and *y* axes on the focal plane, and depicted the results in Fig. S10b (red dotted line and blue dashed line). We note that these two curves match well with each other, implying the good isotropic response of our meta-lens. In addition, we can identify from these two curves that the size of the focal spot is 20 mm, which is labeled by the cyan lines at the full-width-at-half-maximum (FWHM) of the measured patterns. Similarly, we can evaluate the focal length and spot-size of our meta-lens working in reflection space, which are 77 mm and 24 mm, respectively (see Figs. S10c-10d).

## H. Evaluations of the working efficiencies for the full-space meta-lens

In this section, we describe how to evaluate the working efficiency of our meta-lens as studied in Figs. 5-6 of the main text. We first evaluate the working efficiency of the reflective meta-lens under the excitation of an $\hat{x}$-polarized wave. We define the efficiency as the ratio between the power carried by the focal spot and that of the incident beam. Since a direct measurement on the meta-lens' working efficiency is technically impossible, we use a two-step method to approximately evaluate it, following the strategy established in Ref. [24]. The absolute working efficiency can be calculated by $\eta_{foc} = \frac{P_{ref}}{P_{tot}} \times \frac{P_{foc}}{P_{ref}}$. The first term $P_{ref}/P_{tot}$ represents the ratio between the power carried by the totally reflected waves and that of the incident one, while the second one, $P_{foc}/P_{ref}$, is defined by $\frac{P_{foc}}{P_{ref}} = \frac{\oint \vec{S} \cdot \vec{n} \, dA}{\oint \vec{S} \cdot \vec{n} \, dA}$, where $\vec{S}$ is the Poynting vector, $\vec{n}$ is the normal unit vector of the integration area, and the two integrations run over the areas occupied by the focusing spot and that of the meta-lens. The two ratios can both be evaluated in FDTD simulations and experiments, and thus the final absolute efficiencies can be obtained.

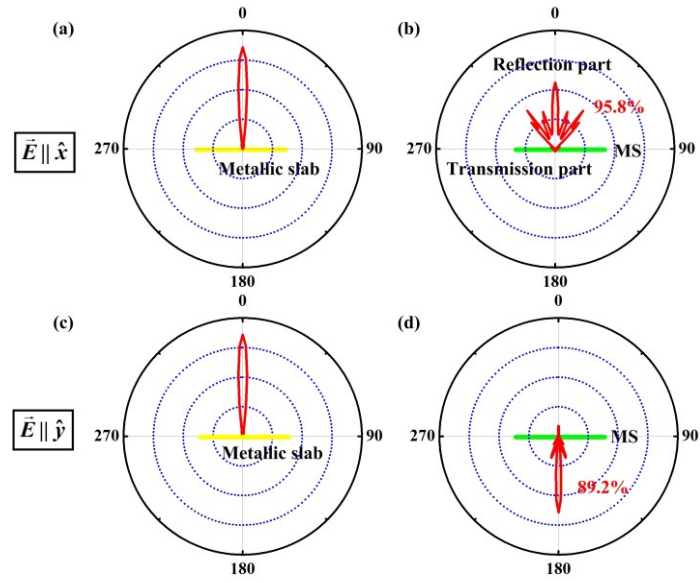

**Figure S11** Measured power distributions of the scattered waves for a metallic slab (a, c) with a same size of the meta-surface and the meta-lens (b, d) illuminated by $\hat{x}$-polarized and $\hat{y}$-polarized incident plane wave at frequency of 10.6 GHz.

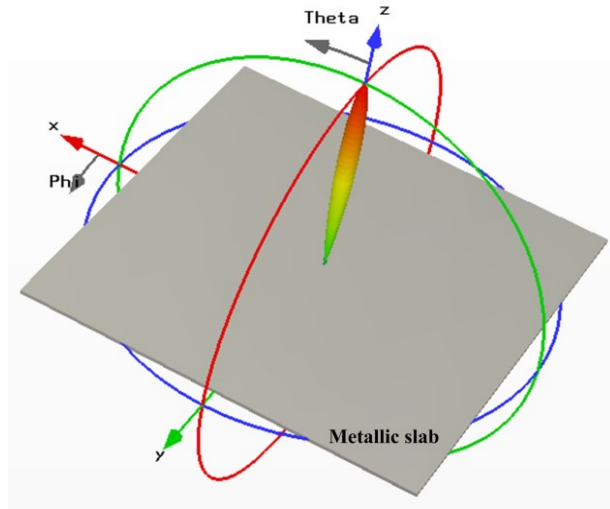

**Figure S12** FDTD simulated 3D power distributions of the scattered waves for a metallic slab illuminated by an $\hat{x}$-polarized incident plane wave at frequency of 10.6 GHz.

We estimate the term $P_{ref}/P_{tot}$ first. For a 3D lens, the reflected/transmitted waves go to all different directions after passing through the focal point. Therefore, we should in principle perform a two-dimensional (2D) integration over the whole solid angle to accurately evaluate the efficiency. However, such a 2D integration is very difficult to perform in experiments. Fortunately, our meta-lens has quite good isotropic responses on the *xy*-plane (see insets to Figs. 6(b) and 6(d) in the man text and Fig. S10b and 10d), and the 3D scattering pattern of a metal slab also exhibit nice isotropic response on the *xy*-plane (see Fig. S12). Therefore, it is a reasonable approximation to take only one principal plane to perform one-dimensional (1D) integrations to obtain both the integrated reflected/transmitted power and the reference value. Shining the meta-lens by a normal-incidence $\hat{x}$-polarized wave at 10.6 GHz, we measured the H-plane scattering pattern and depicted the result in Fig. S11b. As a reference, we then measured the scattering pattern with the metasurface replaced by a metallic slab of the same size, and depicted the scattered-field distributions in Fig. S11a. We then performed integrations over the curves in Fig. S11b and Fig. S11a. The ratio between the two integrated values is found as 95.8%. The missing power is mainly due to dielectric losses and the transmission. We note that this value is also quite close to the averaged value (~0.96) of reflectance of all meta-atoms in our device under $\hat{x}$-polarized excitation, which re-enforced our approximate treatment.

We then evaluated the term $P_{foc}/P_{ref}$. Following the strategy established in Ref. [25], we approximately replaced the formula $\frac{P_{foc}}{P_{ref}} = \frac{\oint}{\oint}$ as $\frac{P_{foc}}{P_{ref}} \approx \frac{\oint}{\oint}$, which is reasonable as our designed meta-lens has a large $F/D$ ratio ($F$: focal length, $D$: lateral size of the lens). Shining the sample with an $\hat{x}$-polarized incident plane wave at $f_0 = 10.6$ GHz, we measured the $|\vec{E}|$ distribution on the focal plane (i.e., the $xy$-plane at $z = -77$ mm) and depicted the results in the inset to Fig. 6b of the main text. We next evaluated the two integrations ($\oint$ $\oint$) based on the measured field pattern, where the first integration runs over the area occupied by the focal spot (defined by the dashed lines in the inset to Fig. 6b) and the second one runs over the entire region occupied by the metasurface. Based on these calculations, we get $\frac{P_{foc}}{P_{ref}} \approx 94.6\%$. Thus, the final working efficiency of our reflective lens is $95.8\% \times 94.6\% \approx 90.6\%$.

We employed the same two-step characterization method to evaluate the working efficiency of our meta-lens working on the transmission mode, based on the formula $\eta_{foc} = \frac{P_{trans}}{P_{tot}} \times \frac{P_{foc}}{P_{trans}}$. The first term $\frac{P_{trans}}{P_{tot}}$ is estimated as 89.2% based on the measured far-field patterns depicted in Fig. S11c-11d, with the missing power carried away by dielectric losses and the reflections. Again, we note that this value is quite close to that (0.88) estimated based on averaging the transmittance of all meta-atoms inside our meta-lens under $\hat{y}$-polarized excitation. The second term $\frac{P_{foc}}{P_{trans}}$ is evaluated as 95.3% based on the measured near-field $|\vec{L}|$ distributions at the focal plane (see inset to Fig. 6d of the main text). Therefore, the product of the two efficiencies gives us the final working efficiency of this functionality, which is roughly 85%.

## I. Phase/amplitude distributions of the full-space bifunctional meta-device studied in Fig. 7

Figure S13 depicts the distributions of reflection/transmission phases ($\varphi_{xx}^r$ and $\varphi_{yy}^t$) and amplitudes ($|r_{xx}|$ and $|t_{yy}|$) of the sample studied in Fig. 7 of the main text, obtained by FDTD simulations. We note that the designed phase distributions match well with the theoretical ones given in Eq. (3), while the reflection/transmission amplitudes of all meta-atoms remain at high values ($|r_{xx}|>0.92$, $|t_{yy}|>0.86$), which guarantees the high efficiencies of our bifunctional meta-device.

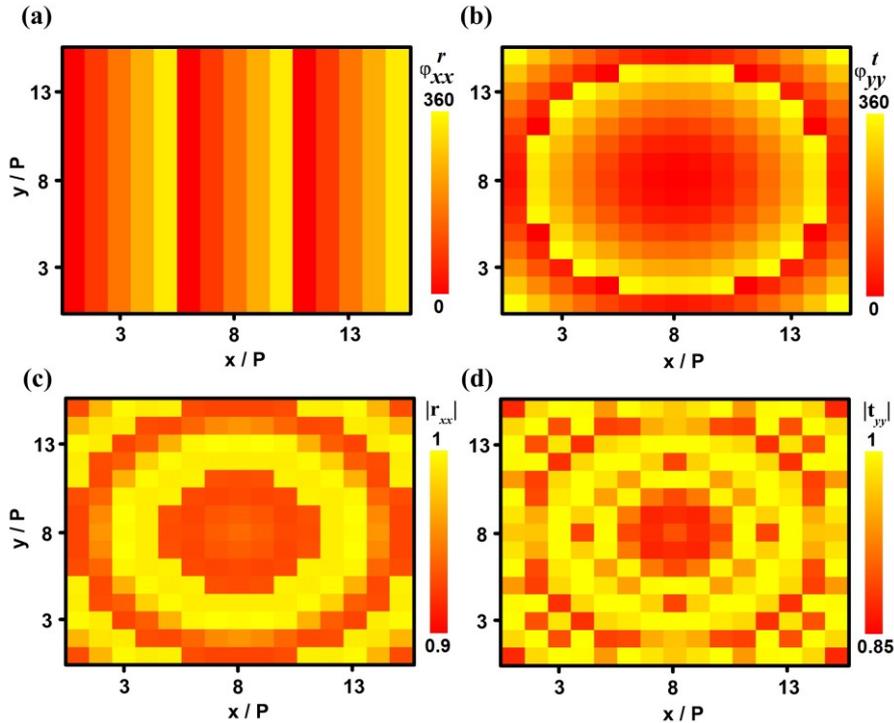

**Figure S13.** FDTD simulated distributions of reflection/transmission phase (a/b) and amplitude (c/d) of the designed meta-device studied in Fig. 7 of the main text, at frequency of 10.6GHz.

## J. Additional simulation/experimental data of the full-space bifunctional meta-device

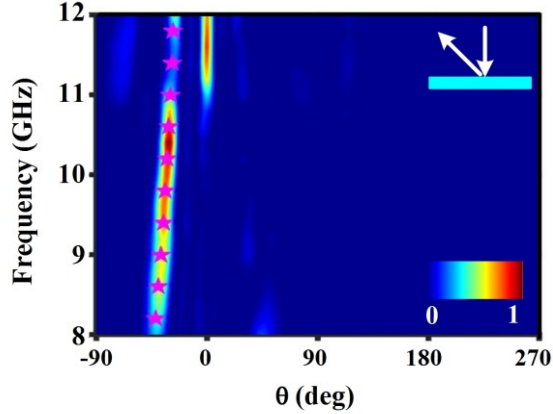

**Figure S14.** FDTD simulated scattered-field intensities (color map) in full space versus frequency and the detecting angles, for our bifunctional meta-surface under the normally incident illumination with polarization of $\vec{E} \parallel \hat{x}$.

Figure S14 contains the FDTD simulation results corresponding to the experimental results shown in Fig. 7. Obviously, they match with each other well.

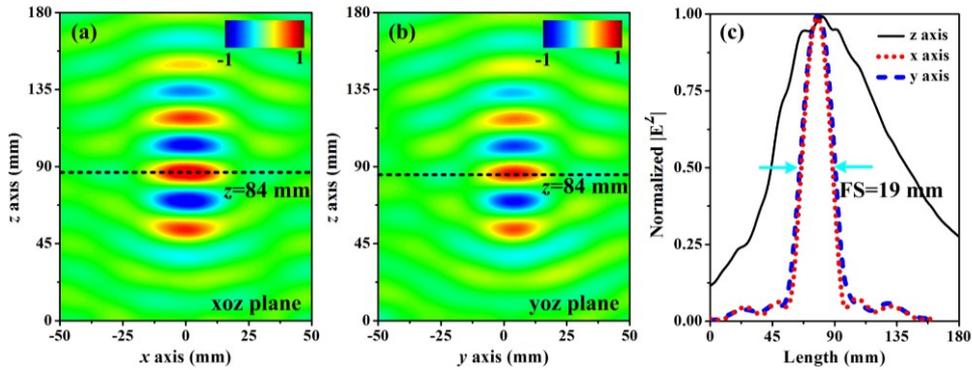

**Figure S15.** Measured $\mathrm{Re}(E_y)$ distributions at *xoz* (a) and *yoz* (b) planes at transmission space for our meta-device illuminated by a $\hat{y}$-polarized wave at 10.6 GHz. (c) Calculated $|E_y|^2$ distributions along three axes from the measured $E_y$ distributions of our meta-lens, respectively.

Figure S15a and S15b show the experimentally measured $\mathrm{Re}(E_y)$ distributions on both *xoz* and *yoz* planes, for the meta-surface (studied in Fig. 7 of the main text) illuminated by a $\hat{y}$-polarized plane wave at 10.6 GHz. To clearly identify the focal point, we depicted in Fig. S15c the measured $|E_y|^2$ distributions along the *z* axis (black line) and found that the focal point is at *z*=84 mm where the $|E_y|^2$ reaches a maximum. Finally, we plotted in Fig. S15c the measured $|E_y|^2$ distributions along *x* and *y* axes on the focal plane at *z*=84 mm (red dot line and blue dash line), from which we can estimate the size of the focusing spot (19 mm) on the focal plane.

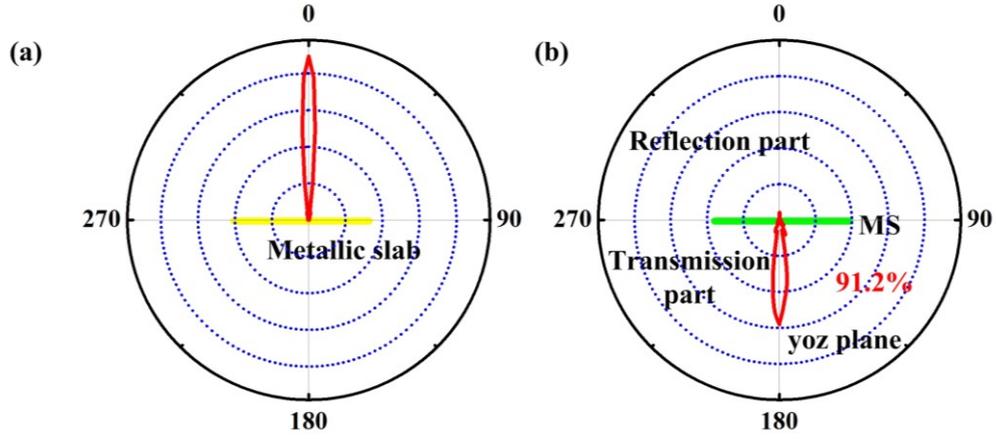

**Figure S16** Measured power distributions of the scattered waves for a metallic slab (a) with a same size of the meta-surface and the meta-lens at (b) *yoz* plane illuminated by $\hat{y}$-polarized incident plane waves at frequency of 10.6GHz.

Finally, we evaluated the working efficiency of the meta-device working as a meta-lens, following exactly the same procedures as described in Section *F*. Figure S16 shows the measured far-field scattering patterns of our bifunctional metasurface and a metal plate of the same size, both illuminated by a normal-incidence $\hat{y}$-polarized microwave at 10.6 GHz. The ration between the integrated values over two curves depicted in Fig. S16 suggested that $\frac{P_{trans}}{P_{tot}} \approx 91.2\%$, which is again in reasonable agreement with the value (0.90) estimated by averaging the transmittance of all meta-atoms inside this meta-device. Based on the measured $|\vec{L}|$ pattern on the focal plane as shown in Fig. 7g of the main text, we further obtained that $\frac{P_{foc}}{P_{trans}} \approx 93.4\%$. The final working efficiency of this functionality is thus 85.2%.